\documentclass[proceedings]{JHEP}
\usepackage{epsfig,amssymb} 
\newcommand{\toline}[1]{}
\newcommand{\phup}{^{\phantom{p}}}

\newcommand{\be}{\begin{equation}}
\newcommand{\ee}{\end{equation}}

\newcommand{\xtra}[1]{{.}}
\renewcommand{\xtra}[1]{{, \tt hep-th/#1.}}

\newcommand{\xtrac}[1]{{.}}
\renewcommand{\xtrac}[1]{{, \tt cond-mat/#1.}}

\newcommand{\mathematica}[1]{{}}

\newcommand{\mm}[2]{{\vphantom{\vbox to 6mm{}}}}
\newcommand{\fract}[2]{{\textstyle\frac{#1}{#2}}}

\newcommand{\CS}{{\cal S}}

\newcommand{\beqcol}{\begin{array}{rcl}}
\newcommand{\eeqcol}{\end{array}}

\newcommand\ZZ{{\mathbb Z}}
\newcommand\CC{{\mathbb C}}
\newcommand\RR{{\mathbb R}}

\newcommand\eq{\begin{equation}}
\newcommand\en{\end{equation}}
\newcommand\bea{\begin{eqnarray}}
\newcommand\eea{\end{eqnarray}}
\newcommand\nn{\nonumber}

\newcommand{\One}{{\hbox{{\rm 1{\hbox to 1.5pt{\hss\rm1}}}}}}
\renewcommand{\One}{{\mathbb 1}}
\renewcommand{\One}{{\rm 1\!\!1}}

\newcommand{\opnup}[1]{\renewcommand{\\}{\\[50 pt]}}

\renewcommand{\tilde}{\widetilde}

\conference{Nonperturbative Quantum Effects 2000}

\title{Ordinary Differential Equations and Integrable Models}

\author{Patrick Dorey, Clare Dunning, Roberto Tateo\\
SPhT Saclay, 91191 Gif-sur-Yvette, France (PED)\\
Dept.~Math.~Sciences, University of Durham, Durham DH1 3LE, UK 
(PED and TCD)\\
UVA, Inst. voor Theoretische Fysica, 1018 XE Amsterdam, The
Netherlands (RT)\\
        E-mails: \email{p.e.dorey@dur.ac.uk},
	\email{tcd1@york.ac.uk}, \email{roberto.tateo@dur.ac.uk}}

\abstract{          
We review a recently-discovered link between
the functional relations approach to
integrable quantum field theories and the properties of certain ordinary
differential equations in the complex domain. 
(Talk given by PED at the TMR conference 
`Nonperturbative~Quantum~Effects~2000')
\centerline{SPhT-T00/136, 
DTP/00187, ITFA 00-18; PRHEP-tmr2000/034, {\tt hep-th/0010148}}\\[-20pt]
}
\begin{document}
\section*{Introduction}
Functional  relations are assuming a growing importance in the study of
integrable lattice models and integrable quantum field theories. The
aim of this talk is to sketch a recently-discovered link between certain
sets of these relations and a rather more classical area of mathematics,
namely the theory of Stokes multipliers and spectral determinants
for ordinary differential equations in the complex domain. For
most of the talk the focus will be on the simplest example of this
`ODE/IM correspondence', connecting $2^{\rm nd}$ order ordinary
differential equations to Bethe ansatz systems of $SU(2)$ type.
However, at the end some recent work extending this to $n^{\rm th}$
order ODEs, and linking them to Bethe ansatz systems associated
with $SU(n)$, will get a mention. 
To the extent that this talk has any logical structure at all, it
is as follows:
\[
\!\!
\left. \begin{array}{ll}
 {\bf (1)\,} & \parbox[t]{2.7cm}{$2^{\rm nd}$ order ODEs\\
	 (Schr\"odinger equations)}\\[32pt]
 {\bf (2)\,} & \parbox[t]{2.8cm}{Functional equations in integrable models}
       \end{array}
\right\}
\to
\begin{array}[t]{ll}
 {\bf (3)} & \parbox{2.2cm}{Connection}\\[8pt]
 &~~~~\downarrow\\[8pt]
 \!\!\!\!{\bf (4)} & \!\!\!\!\parbox{2.35cm}{Generalisations}\\
       \end{array}
\]
Papers directly concerned with this topic include 
\cite{DTa}--\cite{Srev}, but it should be stressed it all 
relies heavily on earlier studies by, among others, Sibuya~\cite{Sha}, 
Voros~\cite{Voros},
and Bender et al~\cite{BT,BB,BBN} (on the ODE side) and by 
Baxter~\cite{Bax}, Kl\"umper, Pearce and
collaborators~\cite{KP,KBP}, Fendley et al~\cite{FLS},
and Bazhanov, Lukyanov and Zamolodchikov~\cite{BLZ1,BLZ2}
on the integrable models side.

\section{Schr\"odinger equations}
Stokes sectors, and their relationship with
eigenvalue problems defined in the complex plane,
will be important
in the following, and to introduce these topics we begin by
describing a class of problems much studied by Bender and collaborators in
recent years. It all began with a question posed by Bessis and
Zinn-Justin, many years ago\dots

\noindent {\bf Question 1:} What does the spectrum of the Hamiltonian
\[
H=p^2+ix^3
\]
look like?

\noindent
This is a cubic oscillator, with purely imaginary coupling $i$. (Strictly
speaking, Bessis and Zinn-Justin, motivated by considerations of the
Yang-Lee edge 
singularity, were interested in more general
Hamiltonians of the form $p^2+x^2+igx^3$, from which the above problem can
be recovered as a strong-coupling limit.)
The
corresponding Schr\"odinger equation is
\[
-\frac{d^2}{dx^2}\psi(x)+ix^3\psi(x)=E\psi(x)
\]
and we will initially say that 
the (possibly complex) number $E$ will be in the spectrum if and only
if, for that value of $E$, the equation has a solution $\psi(x)$ on the
real axis which decays both at $x\to-\infty$ and at 
$x\to+\infty$:\footnote{To be more precise, the decay should be fast
enough that $\psi$ lies in $L^2(\RR)$, the space of square-integrable
functions. This means that we are actually discussing the so-called {\em
point spectrum} of $H$ -- see, for example, \cite{REL}.}
\[
\epsfxsize=.70\linewidth
\epsfbox{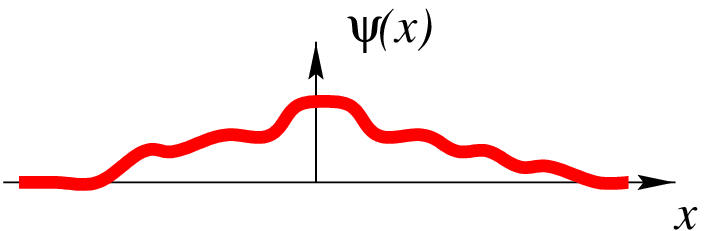}
\]
Note that the wavefunction $\psi(x)$ will inevitably be complex.
Since the Hamiltonian is not (at least in any obvious way) Hermitian, 
the usual arguments to show that all of the eigenvalues $E$
must be real do not apply. 
Nevertheless, perturbative and numerical studies led
Bessis and Zinn-Justin to the following {\bf conjecture}:

\noindent
$\bullet$ the spectrum of $H$ \underline{is} real, and positive.

In 1997 Bender and Boettcher~\cite{BB} proposed a nice generalisation of this
problem:

\noindent {\bf Question 2:} What is the spectrum of
\[
H=p^2-(ix)^N\qquad\quad\mbox{($N$ real, $>0$)}\,?
\]
Later, it will turn out that the passage from question 1 to question 2
corresponds to a change in a coupling constant in a sine-Gordon model, or
of a quantum group deformation parameter in a Bethe ansatz system. But for
now, the generalisation is appealing because it unites 
into a single family of eigenvalue problems both
the $N{=}3$ case, for
which we have the Bessis-Zinn-Justin conjecture, and
the more easily-understood $N{=}2$ case, the harmonic oscillator.
The Schr\"odinger equation is now
\[
-\frac{d^2}{dx^2}\psi(x)-(ix)^N\psi(x)=E\psi(x)
\]
and, as before, we look for those values of $E$ at which there
is a solution along the real $x$-axis which decays at both plus and minus
infinity. Two details need extra care: for non-integer values of $N$, the
`potential' $-(ix)^N$ is not single-valued; and when $N$ hits
$4$, the naive definition of the eigenvalue problem runs
into difficulties. The first problem is easily cured by adding a
branch cut along the positive imaginary $x$-axis, but the second is 
more subtle and will be discussed in greater detail below. 

This caveat aside, there is already a surprise while $N$ remains
below $4$. Figure~\ref{fig1} is taken from \cite{DTb}, and it
reproduces the results of \cite{BB}.
Ignoring for a moment the region $N>4$, it is clear 
that something strange occurs as $N$ decreases through $2$ -- 
infinitely-many eigenvalues pair off and become complex,
and only finitely-many remain real. By the time $N$ reaches $1.5$,
all but three have become complex, and as $N$ tends to $1$ the last real
eigenvalue diverges to infinity. In fact, at $N{=}1$ the problem has no
eigenvalues at all, as can be seen by solving the relevant
Schr\"odinger equation in terms of an Airy function.
For $N>2$, the numerically-obtained spectrum is entirely real, and 
positive, and so the conjecture of Bessis and Zinn-Justin has found a 
natural  generalisation. The `phase transition' 
to infinitely-many complex
eigenvalues at $N{=}2$ was interpreted in \cite{BB} as a spontaneous
breaking of ${\cal PT}$ symmetry.

\[
\begin{array}{c}
\refstepcounter{figure}
\label{fig1}
\epsfxsize=1.0\linewidth
\epsfysize=0.8\linewidth
\epsfbox{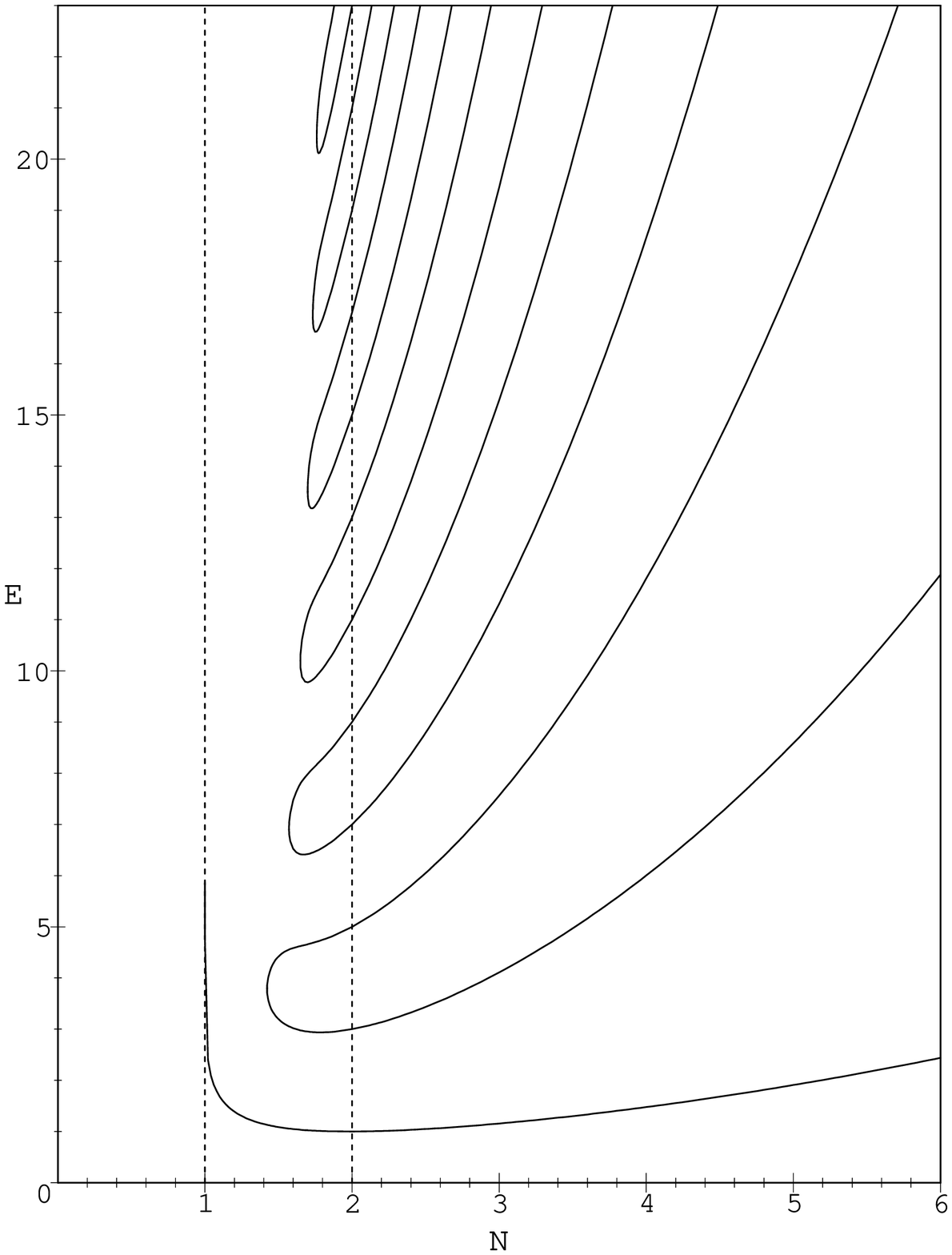}~~~
\\[-20pt]
\parbox{0.92\linewidth}{
\raggedright
Figure \ref{fig1}: 
$H=p^2-(ix)^N$\,:\\
{}~~~real eigenvalues as
a function of $N$
}~~~
\end{array}
\]

Although figure \ref{fig1} agrees with the plot in~\cite{BB}, 
it was obtained in \cite{DTb} by
an entirely different route -- rather than making a direct numerical attack
on the ordinary differential equation, a nonlinear integral equation for
the relevant spectral determinant was solved. This method of solving
such eigenvalue problems is a byproduct of the ODE/IM correspondence 
and appears to be new, though it owes a heavy debt to
earlier work of Voros~\cite{Voros}. Numerically, it is rather efficient --
see for example 
the tables in \cite{DTa} of eigenvalues of various anharmonic oscillators.

Another idea motivated by
the correspondence 
is the notion~\cite{DTb} to study the effect of an additional
angular-momentum term
$l(l{+}1)x^{-2}$ 
on the Bender-Boettcher problem. For $-1<l<0$, this 
turns out to have a
remarkable effect on the behaviour of the spectrum as the
$N{=}2$ phase transition is crossed.
\[
\begin{array}{c}
\refstepcounter{figure}
\label{fig2}
\epsfxsize=0.72\linewidth
\epsfysize=0.6\linewidth
\epsfbox{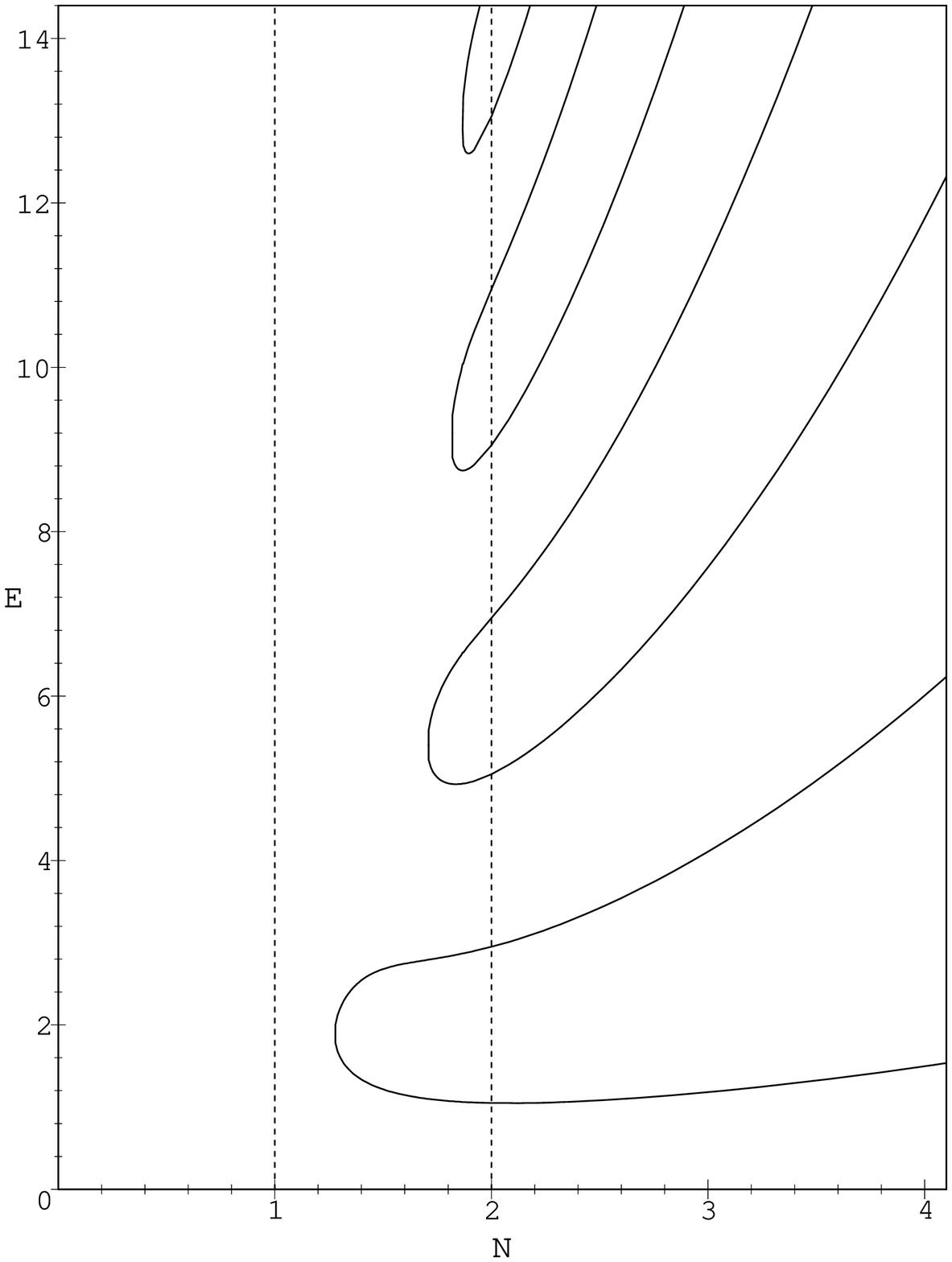}~~~
\\[-10pt]
\parbox{.92\linewidth}{\raggedright
Figure \ref{fig2}: 
$H=p^2-(ix)^N-0.024735\,x^{-2}$\,:\\
{}~~~real eigenvalues
as a function of $N$
}
\end{array}
\]

\noindent
Figure \ref{fig2} zooms in on this part of the
spectral plot for $l=-0.025$, and reveals
that the picture has changed dramatically -- 
the connectivity
of the real eigenvalues has been completely reversed, so that while for
$l{=}0$ (the original Bender-Boettcher problem) the first and second
excited states pair off, at $l=-0.025$ 
the first excited state
is instead paired with 
the ground state,
and so on up the spectrum. 
With this in mind, it may be a little
hard to see how
it is possible to pass between the sets of spectra depicted in figures
\ref{fig1} and \ref{fig2} simply by 
varying the continuous parameter $l$ from zero to $-0.025$. 
The puzzled reader is invited to have a look at figure~2 of \cite{DTb} to
resolve the mystery.

There remains one piece of unfinished business: what goes wrong at
$N{=}4$, and what can be done to resolve it? On figures \ref{fig1}
and \ref{fig2}, the
curves continue smoothly past $N{=}4$, but in fact this is only
achieved by implementing a suitable distortion of the problem as
originally posed. Consider the situation precisely at $N{=}4$\,:
the Hamiltonian is $p^2-x^4$, an `upside-down' quartic oscillator, and a
simple WKB analysis (about which more shortly) shows, instead of 
the exponential growth or decay more generally found, wavefunctions
behaving as $x^{-1}\exp(\pm ix^3/3)$ as $x$ tends to plus or minus infinity. 
{\em All} solutions thus decay, albeit algebraically, and this 
complicates matters significantly. The problem moves from what is called
the limit-point to the limit-circle case (again, see \cite{REL}), 
and additional boundary conditions should be imposed at infinity if
the spectrum is to be discrete.  
While interesting in its own right, this is clearly
not the right eigenproblem 
if we wish to find a smooth continuation from the region $N<4$.
Instead, it is more fruitful to enlarge the perspective and
treat $x$ as a genuinely complex variable. This 
has been discussed by many authors, and is particularly emphasised in
the book by Sibuya \cite{Sha}; 
the treatment which follows is very close to that of
\cite{BT,BB}.

The key is to examine the behaviour of solutions as $|x|\to\infty$ along a
general ray in the complex plane, even though the only two rays that we
initially need are the positive and negative real axes.
The WKB approximation tells us that
\[
\psi(x)\sim P(x)^{-1/4}\,e^{\pm\int^x\!\sqrt{P(t)}dt}
\]
as $|x|\to\infty$, with $P(x)=-(ix)^N-E$. (This is easily derived by
substituting $\psi(x)=f(x)e^{g(x)}$ into the ODE.) 
Since we set the problem up with a branch cut running up the
positive-imaginary axis, it is natural to define a ray in the complex
plane by setting $x=\rho e^{i\theta}\!/i$ with $\rho$ real:
\[
\epsfxsize=.55\linewidth
\epsfbox{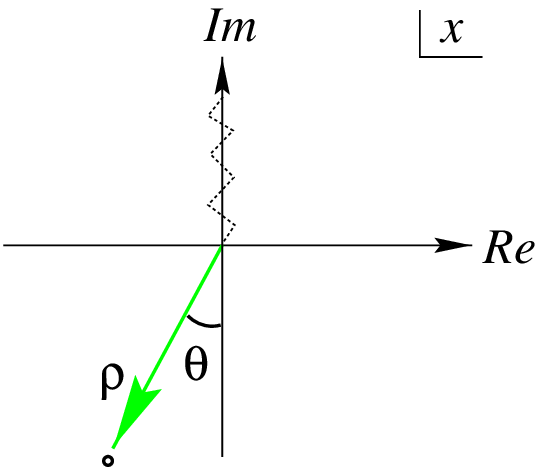}
\]

\noindent
For $N>2$, the
asymptotic is not changed if $P(x)$ is replaced by $-(ix)^N$, and
substituting into the general formula we see two possible
behaviours, as expected of a second-order ODE:
\[
\psi_{\pm}\sim P^{-1/4}\exp\left[\pm\fract{2}{N{+}2}e^{i\theta(1{+}N/2)}
\rho^{1{+}N/2}\right]\,.
\]
For most values of $\theta$, one of these solutions
will
be exponentially growing, the other exponentially decaying. But whenever
$\Re e[e^{i\theta(1{+}N/2)}]=0$, the two solutions swap roles and there is
a moment when both oscillate, and neither dominates the other. 
The relevant values of $\theta$ are
\[
\theta=
\pm\frac{\pi}{N{+}2}~,~
\pm\frac{3\pi}{N{+}2}~,~
\pm\frac{5\pi}{N{+}2}~,~\dots
\]
(Confusingly, the rays that these values of $\theta$ define
are sometimes called `anti-Stokes lines', and sometimes `Stokes
lines'.)
Whenever one of these lines lies along the positive
or negative real axis, the eigenvalue
problem as originally stated becomes much more delicate, for the
reasons described above.
Increasing $N$ from $2$, the first time that this happens
is $N{=}4$, the case of the
upside-down quartic potential. But now we see that the
problem is easy to avert -- it arose because the line along which the
wavefunction was being considered, namely the real axis, happened to
coincide with an anti-Stokes line\footnote{as just mentioned, some 
would call this a
Stokes line.}. But since all functions
involved are analytic, there is nothing to stop us from examining
the wavefunction along some other contour in the complex plane.  In
particular, before $N$ reaches $4$, the two ends of the contour can be bent
downwards from the real axis without changing the spectrum, so long as
their asymptotic directions do not cross any anti-Stokes lines in the
process. 
Having thus distorted the original problem, $N$ can be increased
through $4$ without any difficulties.
The situation for $N$ just bigger than $4$ is illustrated below, 
with the anti-Stokes lines shown dashed and the wiggly line a
curve along which the wavefunction $\psi(x)$ can be
defined.

\[
\epsfxsize=.55\linewidth
\epsfbox{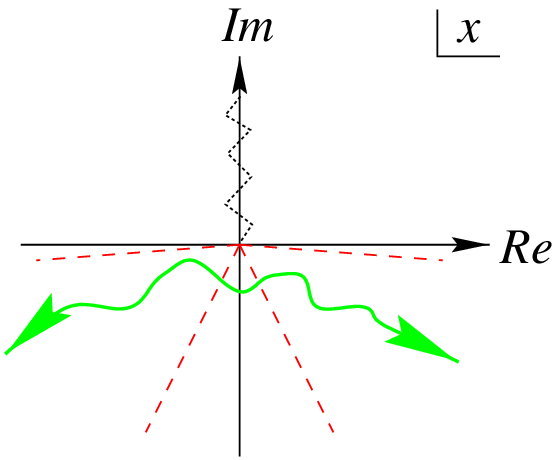}
\]
The wedges between the dashed lines 
are called {\em Stokes
sectors}, and in directions out to infinity which lie inside these
sectors, wavefunctions either grow or decay exponentially, leading to 
eigenvalue problems with straightforward, and discrete, spectra. 
Note that once $N$ has passed through $4$, as in the figure, the real axis
is once again a `good' quantisation contour -- but for a {\em different}
eigenvalue problem, which is {\em not} the analytic continuation of the 
original $N<4$ problem to that value of $N$. (For the analogue of
figure~\ref{fig1} for this new problem, see figure~20 of \cite{BBN}.)

There is a lesson to be drawn from all of
this~\cite{Sha,BT,BB,BBN}. Associated with an ODE
of the type under consideration there
are many different
eigenvalue problems, each defined 
by specifying a {\em pair} of Stokes sectors,
and then asking for the values of $E$ at which there exist solutions  
to the equation which 
decay exponentially 
in both simultaneously. For a given value of $N$,
the two sectors which cover the positive and negative real axes may appear
to be the most natural choice, but if we want to discuss analytic
continuation then all must be put on an equal footing. 
This picture will find a precise analogue on the integrable models
side of the correspondence, but before describing this we need to review
some more basic material.

\section{Functional relations in integrable models}
In this section a very rough caricature of the `functional relations'
approach to integrable models will be given. A number of other speakers at
the conference talked about this topic, in particular J.-M.~Maillet, 
R.~Poghossian and F.~Smirnov, and their contributions should be consulted
for more in-depth reviews. Not to forget, of course, the book~\cite{Bax} by
Baxter\dots

We will discuss the six-vertex model, defined initially
on an $N\times M$ lattice,
with periodic boundary conditions and $N/2$ even. 
On each (horizontal or vertical) link of
the lattice, we place a spin $1$ or $2$, conveniently depicted by an arrow
pointing either left or right (for the horizontal links) or up or down
(for the vertical links). Only those configurations of spins
which preserve the `flux' of arrows through each vertex are permitted. Locally
this gives just six options (hence the name of the model) to which
Boltzmann weights are assigned as follows:
\[
R^{11}_{11}=R^{22}_{22}=a
\]
\[
R^{21}_{12}=R^{12}_{21}=b
\]
\[
R^{12}_{12}=R^{21}_{21}=c
\]
The relative probability of finding any given configuration is found by
multiplying together the Botzmann weights for the individual vertices, and
a first quantity to calculate is just the sum of these numbers over all
possible configurations -- the partition function, $Z$. Very crudely speaking,
a model is said to be integrable if it is possible to evaluate quantities
such as $Z$ (or even better, the free energy)
exactly, at least in the limit where $N$ and $M$ both tend to
infinity. The model under discussion turns out to be integrable in this
sense for all values of $a$, $b$ and $c$. The overall normalisation
factors out trivially from all calculations, and it is convenient to
parametrise the remaining two degrees of freedom 
using a pair of variables $\nu$ (the spectral parameter) and $\eta$
(the anisotropy):
\[
a=\sinh(\nu-i\eta)~,~~b=\sinh(\nu+i\eta)~,~~c=\sinh(2i\eta)~.
\]
To calculate $Z$, one line of attack proceeds via the so-called {\em
transfer matrix}, $T$:
\[
T^{\alpha_1'\alpha_2'\phantom{q}\!\!\!\dots\alpha'_N}%
_{\alpha_1\alpha_2\phantom{'}\!\!\dots\alpha_N}
{}~{}={}~{}
\sum_{\{\beta_i\}}
R^{\alpha_1'\beta_2}_{\beta_1\alpha_1}
R^{\alpha_2'\beta_3}_{\beta_2\alpha_2}
\dots
R^{\alpha_N'\beta_1}_{\beta_N\alpha_N}
\]
The job of $T$, a $2^N\times 2^N$ matrix,
is to perform the sum over a set of horizontal links. 
In this picture
the indices of $T$ correspond to the spin
variables sitting on the vertical links, which
can now be summed by
matrix multiplication. Thus:
\[
Z=\mbox{Trace}\left[T^M\right]~.
\]
The next step is to compute via a diagonalisation of $T$. For example,
the free energy per site in the limit $M\to\infty$ can be obtained as
\[
f=\fract{1}{N\!M}\log Z=
  \fract{1}{N\!M}\log \mbox{Trace}\left[T^M\right]\sim\fract{1}{N}\log t_0
\]
where $t_0$ is the largest eigenvalue of $T$ (corresponding to the ground
state). Note that the eigenvalues $t_0$, $t_1$\,\dots are all
functions of $\nu$ and $\eta$. However, there is still work to be done
to find out what 
these functions are. At this point we just state that
there exists a technique, the (algebraic) Bethe ansatz, for doing
this. Skipping all details, the method works in two stages:

\noindent
(i) \underline{Guess} a form for an eigenvector of $T$, depending on a
finite number of parameters $\nu_1,\dots \nu_n$ (the 
`{\em roots}').

\noindent
(ii) Discover that this guess only works if the $\{\nu_i\}$ together
solve a certain set of coupled equations (the `{\em Bethe ansatz
equations}').

These equations will be written down shortly, but first we describe a
particularly neat reformulation that was found by Baxter. 
The first input is the fact that the transfer matrices commute at
different values of $\nu$:
\[
[T(\nu),T(\nu')]=0\,.
\]
This means that they can be diagonalised simultaneously, with
$\nu$-independent eigenvectors, and it allows us to focus on the
individual eigenvalues $t_0(\nu)$, $t_1(\nu)$,\dots as functions
of $\nu$. From the explicit form of the Boltzmann weights
these functions are entire, and $i\pi$-periodic.

Now for the key result: for each eigenvalue function $t(\nu)$,
there exists an auxiliary function $q(\nu)$, also entire and
(at least for the ground state)
$i\pi$-periodic, such that
\[
t(\nu)q(\nu)=a^Nq(\nu+2i\eta)+b^Nq(\nu-2i\eta)\,.
\]
We shall call this the T-Q relation, though this phrase might more
properly
be reserved for the corresponding matricial equation, involving
$T(\nu)$ and another matrix $Q(\nu)$, from which the above
can be extracted when acting on eigenvectors. It is not immediately
clear why this result represents progress -- we started with one unknown
function $t(\nu)$, and have now been told that if we multiply this by
another unknown function $q(\nu)$, we recover two copies of
that same function at shifted
values of its argument. However, the fact that both $t$ and $q$ are
entire makes this condition much more restrictive than might first
appear.

To make the match with the ODEs described in the last section, we will
need to take a certain large-$N$ limit, simultaneously shifting $\nu$
and rescaling the TQ relation. This has
the effect of eliminating the factors $a^N$ and $b^N$, and since it
also simplifies the formulae, from here on we will assume that this has
been done.
The relation becomes
\[
t(\nu)q(\nu)=q(\nu+2i\eta)+q(\nu-2i\eta)\qquad
\mbox{(TQ)}\!\!\!\!\!\!
\]
and
the $i\pi$-periodic function
$q(\nu)$ can be written
as a product over its zero positions as
\[
q(\nu)=\prod_l\sinh(\nu-\nu_l)~.
\quad\qquad\qquad~
\mbox{(Q)}
\!\!\!\!\!\!
\!\!\!\!\!\!
\!\!\!\!\!\!
\]
Strictly speaking this product only converges if the number of zeroes
is finite, which is not true in the limit we consider --
the $\nu_l$ accumulate
at infinity. Given certain growth conditions, $q(\nu)$
can more correctly be written as
a product of factors
$(1-e^{2(\nu{-}\nu_l)})$\,.
We are only interested in giving the flavour of the argument here,
so having mentioned this caveat
we retain the form appropriate for finite $N$.
Now the reasoning goes as follows. First, we know from (TQ) that $t$
is fixed by $q$, and from (Q) that $q$ is fixed by the set
$\{\nu_i\}$. To fix the $\{\nu_i\}$, 
set $\nu=\nu_i$ in (TQ). On the LHS we then have
$t(\nu_i)$, which is nonsingular since $t$ is entire, times
$q(\nu_i)$ which is zero by (Q). Thus the LHS vanishes, and
rearranging we have
\[
\frac{q(\nu_i+2i\eta)}{q(\nu_i-2i\eta)}=-1~,\qquad i=1,2\dots
\]
or, using (Q) one more time,
\[
\prod_l\frac{\sinh(\nu_i-\nu_l+2i\eta)}%
{\sinh(\nu_i-\nu_l-2i\eta)}=-1~,\quad i=1,2\dots
\]
This is exactly the Bethe ansatz equation (BAE) for the problem, 
with
the $\nu_i$ the roots. The formula for $t(\nu)$ implied by
(TQ) then matches that resulting from a direct application of the
algebraic Bethe ansatz. It is important to realise
that the BAE does not have a unique
solution, but a discrete set of them
(infinite in the $N\to\infty$ limit),
matching the fact $T$ has many eigenvalues\footnote{We won't go into the
question of the completeness of the BAE solutions
here; see~\cite{Barry} for a recent discussion.}.
To select a particular solution, supplementary analyticity conditions
should be imposed. In particular, the ground state emerges if we
require that all of the $\nu_i$ lie on the real axis.

So far we have been discussing the behaviour of lattice models. However,
Bazhanov, Lukyanov and Zamolodchikov were able
to construct analogues of the $T$ and $Q$ operators directly in the
context of a continuum quantum field theory~\cite{BLZ2}, using a
free-field representation of the
massless limit of the sine-Gordon model. The functional relation (TQ)
is then most usually written in terms of a variable
$\lambda$, on which the `shifts' on the RHS
act multiplicatively, as follows:
\bea
&&\!\!\!\!\!\!\!\! T(\lambda)A_{\pm}(\lambda)={}\nn\\[3pt]
&&\qquad e^{\mp 2\pi ip}A_{\pm}(q^{-1}\lambda)+
e^{\pm 2\pi ip}A_{\pm}(q\lambda)
\quad\mbox{(TQ$'$)}
\!\!\!\!\nn
\eea
Here $T$ and $A_{\pm}$ are entire functions of $\lambda^2$,
$q=e^{i\pi\beta^2}$ with $\beta$ the sine-Gordon coupling, and $p$, an
extra parameter compared to the previous discussion, is
related to the possibility of adding a twist to the periodic boundary
conditions (an option which also exists on the lattice). Note also that 
$q$ can be interpreted as a quantum group deformation parameter.

\section{The TQ/ODE connection}
The goal now is to show that (TQ$'$) also arises naturally in connection
with the eigenvalue problems discussed in section 1. First, we 
need to develop our treatment of ordinary differential equations in
the complex domain a little further, relying largely on the
book by Sibuya~\cite{Sha}.

\noindent
Consider the ODE
\[
\left[
-\frac{d^2}{dx^2}+P(x)\right]\,\psi(x)=0
\qquad\qquad\mbox{(*)}
\!\!\!\!\!\!\!\!\!\!\!\!
\!\!\!\!\!\!\!\!\!\!\!\!
\!\!\!\!
\]
where $P(x)=x^{2M}-E$, and $M$ is real and positive. 
(This is the Bender-Boettcher problem with
$N=2M$, $x\to x/i$ and $E\to -E$ -- a change which is made purely for
convenience.) Then~\cite{HseihSha}
the ODE (*) has a solution $y(x,E)$ such that

\noindent
{\bf (i)} $y$ is an entire function of $(x,E)$ 

\noindent
{}~~~~{\small [\,though 
$x$ lives on a cover of $\CC \backslash \{0\}$
if $2M{\notin} \ZZ$ \,]}

\noindent
{\bf (ii)} as $|x|\to\infty$ with $|\arg\, x|<3\pi/(2M{+}2)$,
\bea
y\,&\sim& ~x^{-M/2}\exp\left[-\fract{1}{M{+}1}\,x^{M{+}1}\right]\nn\\
y'&\sim& -x^{M/2}\exp\left[-\fract{1}{M{+}1}\,x^{M{+}1}\right]\,\nn
\eea
{}~~~~{\small [\,though there are small modifications for $M\leq 1$\,]}\\
These properties fix $y$ uniquely; to understand where they come from we 
quickly recall the discussion of section 1. With the shift from $x$ to $x/i$, 
the anti-Stokes lines for the current problem are
\[
\arg(x)=
\pm\frac{\pi}{N{+}2}~,~
\pm\frac{3\pi}{N{+}2}~,~
\dots
\]
and in between them lie the Stokes sectors, which we label by defining
\[
\CS_k=\left|\arg(x)-
\frac{2\pi k}{2M{+}2}\right|<\frac{\pi}{2M{+}2}\,.
\]
The asymptotic quoted in property {\bf (ii)} is just the WKB result
in $\CS_{-1}\cup\CS_0\cup\CS_1$\,. One more piece of notation: an 
exponentially-growing solution in a given sector is called {\em dominant}
(in that sector); one which decays is called {\em subdominant}. It is
easy to check that $y$ as defined above is subdominant in $\CS_0$, and dominant
in $\CS_{-1}$ and $\CS_1$. Note that subdominant solutions to a second-order
ODE are unique up to a constant multiple; this is why the quoted asymptotics
pin down $y$ uniquely.

Having identified one solution to the ODE, we can now generate
a whole family using a trick due to Sibuya. Consider the function
$
\hat y(x,E)=y(ax,E)
$
for some (fixed) $a\in\CC$. From (*),
\[
\left[
-\frac{d^2}{dx^2}+a^{2M+2}x^{2M}-a^2E\right]\,\hat y(x,E)=0\,.
\]
(This is sometimes given the rather-grand name of
 `Symanzik rescaling'.)
If $a^{2M+2}{=}1$, it follows that
$\hat y(x,a^{-2}E)$ solves (*).
Setting 
\[
\omega=e^{2\pi i/(2M{+}2)}
\]
and
\[
y_k(x,E)=\omega^{k/2}y(\omega^{-k}x,\omega^{2k}E)
\]
we therefore have the key statements

\noindent
$\bullet$ $y_k$ solves (*) for all $k\in\ZZ$\,;

\noindent
$\bullet$ up to a constant, $y_k$ is the unique solution to (*) 
subdominant in $\CS_k$.
{}~~{\small [\,This follows easily via the asymptotic of $y$.\,]}

\noindent
$\bullet$ each pair $\{y_k,y_{k+1}\}$ forms a {\em basis} of solutions for 
(*).
{}~~{\small [\,This follows on comparing the asymptotics of $y_k$ and $y_{k+1}$
in either $\CS_k$ or $\CS_{k+1}$.\,]}

We have almost arrived at the T-Q relation. First, expand $y_{-1}$ in the
$\{y_0,y_1\}$ basis:
\[
y_{-1}(x,E)=C(E)y_0(x,E)+\widetilde C(E)y_1(x,E)~.
\]
We will call this a {\em Stokes relation}, with
the coefficients $C(E)$ and $\widetilde C(E)$ {\em Stokes
multipliers}. They can be expressed in terms of Wronskians. A quick
reminder~\cite{CL}: the {\em Wronskian} of two functions $f$ and $g$ is
\[
W[f,g]=fg'-f'g\,.
\]
For two solutions of a second-order ODE with vanishing first-derivative
term, $W[f,g]$ is independent of $x$, and vanishes if and only if
 $f$ and $g$
are proportional. To save ink we set
\[
W_{k_1,k_2}=W[y_{k_1},y_{k_2}]
\]
and record the following two useful properties:
\[
W_{k_1+1,k_2+1}(E)=W_{k_1,k_2}(\omega^2E)~,~~ W_{0,1}(E)=2i\,.
\]
Now by `taking Wronskians' of the Stokes relation first
with $y_1$ and then with $y_0$
we find
\[
C=\frac{W_{-1,1}}{W_{0,1}}~~,\quad\tilde C=-\frac{W_{-1,0}}{W_{0,1}}=-1
\]
and so the relation can be rewritten as
\[
C(E)y_0(x,E)=y_{-1}(x,E)+y_1(x,E)~,
\]
or, in terms of the original function $y$, as
\bea
&&C(E)y(x,E) \nn\\
&&\qquad =\omega^{-1/2}y(\omega x,\omega^{-2}E)+
\omega^{1/2}y(\omega^{-1} x,\omega^2E)\,.\nn
\eea
This looks very like the T-Q relation! The only fly in the ointment is
the 
$x$-dependence of the function $y$. But this is easily fixed:
just set $x$ to zero. We can also take a derivative with respect to
$x$ before setting it to zero, which swaps
the phase factors $\omega^{\pm1/2}$. So we define
\[
D^-(E)=y(0,E)~~,\quad D^+(E)=y'(0,E)\,.
\]
(The notation will be justified shortly.) Then the Stokes relation
implies
\bea
&&\!\!\!\!\!\!\!\! 
C(E)D^{\mp}(E)={} \nn\\[3pt]
&&\quad
\omega^{\mp 1/2}D^{\mp}(\omega^{-2}\!E)+
\omega^{\pm 1/2}D^{\mp}(\omega^2\!E)
\quad\mbox{(CD)}
\!\!\!\!\nn
\eea
Finally we are ready to make the comparison. If we set 
\[
\beta^2=\frac{1}{M{+}1}~~,\qquad
p=\frac{1}{4M{+}4}
\]
then the match between (TQ$'$) and (CD)
is perfect, with the following correspondences between objects
from the IM and ODE worlds:
\bea
T~&\leftrightarrow& ~C\nn\\
A_{\pm}~&\leftrightarrow &~D_{\mp}\nn
\eea
How should we think about $C$ and $D$? In fact they are spectral determinants.
Recall that $C(E)$ is proportional to $W_{-1,1}(E)$. 
Thus $C(E)$ vanishes if and only if
$W[y_{-1},y_1]=0$, in other words if and only if
$E$ is such that $y_{-1}$ and $y_1$ are linearly dependent. But this means that
(*) has a solution decaying in the two sectors $\CS_{-1}$ and $\CS_1$
simultaneously, which is exactly the spectral problem
discussed in section~1, modulo the trivial redefinitions of $x$ and $E$.
This is enough to deduce that, up to a factor of an entire function with
no zeroes, $C(E)$ is the spectral determinant for the Bender-Boettcher
problem. Even this ambiguity can be eliminated,
via Hadamard's factorisation theorem,
once the growth properties of the functions involved have been checked;
see \cite{DTb} for details. To see that the functions $D^{\pm}$ are spectral
determinants is even more easy: first, we note that by their very definition
the functions $y(x,E)$ decay at (real) $x\to\infty$ for all values of $E$.
If $D^-(E)=y(0,E)=0$, then this solution, decaying at $+\infty$, also vanishes 
at $x=0$, while if $D^+(E)=0$, then it has vanishing first derivative
there. A moment's thought shows that this corresponds to there existing odd
or even, respectively, wavefunctions for the equation on the full real axis
with potential $|x|^{2M}$. (It was for this reason that the functions $D^{\pm}$
were so labelled.)

This insight allows us to fill in one gap in the correspondence. While the 
T-Q relation is very restrictive, as remarked in section~2 it does not have
a unique solution. So to say that $D^-(E)$ is `equal' to $A_+(\lambda)$
begs the question: which $A_+(\lambda)$? 
To answer, we first note that,
in contrast to the Bender-Boettcher problem, the full-line problem
with $|x|^{2M}$ potential (or equivalently, the half-line problem with
$y(0)=0$ boundary conditions) {\em is} self-adjoint, and so all of its 
eigenvalues are real. Back in the integrable model, 
the only solution to the BAE
with all roots real is known to be the ground state, so the question is
answered: the relevant $A_+(\lambda)$ is that corresponding to the ground
state of the model.

As it stands, the correspondence is still
not entirely satisfactory since, for
each value of $\beta^2$, it picks out just one value of 
$p$. A more complete mapping would find partners for the BAE at other values
of the twist parameter
as well. This was sorted out very shortly after the original
observation of the correspondence in \cite{DTa}: in \cite{BLZa}, Bazhanov,
Lukyanov and Zamolodchikov pointed out that the ODE (*)
should be generalised
to
\[
\left[
-\frac{d^2}{dx^2}+x^{2M} +\frac{l(l{+}1)}{x^2}-E\right]\psi(x)=0\,.
\]
(This observation, combined with the discovery of the role of the 
$T$ operator made in \cite{DTb}, provided the motivation to study the
spectra shown in figure \ref{fig2} of section 1.)
The previous mapping between parameters becomes
\[
\beta^2=\frac{1}{M{+}1}~~,\qquad
p=\frac{2l{+}1}{4M{+}4}\,,
\]
and varying $l$ away from zero allows us to explore the other values 
of $p$.  This is a continuation through continuous values
of angular momentum in a  radial (three-dimensional) Schr\"odinger
equation -- in other words, non-relativistic Regge theory! A little more 
care is needed in the definition of $D^{\pm}$ once $l(l{+}1)$ is
nonzero, since the equation acquires a regular 
singularity at the origin. The resolution is to
match the solutions $y_k$ onto solutions $\psi_{\pm}$ with simple scaling
behaviours at the origin; the details can be found in \cite{BLZa,DTb}.

Two more points deserve a mention. First, studies of integrable models
had already shown how to transform a T-Q relation into a nonlinear integral 
equation (NLIE), which in turn can be solved by numerical iteration rather
easily~\cite{KBP,DDV,BLZ2}. The NLIE is particularly simple for the ground
state, and it was this that allowed the spectral plots of section~1 to be
obtained in \cite{DTb} with relatively little pain, building on the
checks for specific cases performed in \cite{DTa}.
Second, we should mention that there is another strand to the functional
relations approach to integrable models, based on the so-called fusion 
hierarchy and its truncations (see for example \cite{KP,BLZ1}). This proceeds
via the definition of fused transfer matrices $T_j$, $j=0$,
$\frac{1}{2}$, $1$,
$\frac{3}{2}$, \dots (with $T_0=1$ and the original $T$ identified 
with $T_{1/2}$), and ultimately leads to another set of nonlinear integral
equations, often referred to as being of `TBA type'~\cite{ZamTBA}.
Obviously it would be nice to find a role for these objects as well, and
it turns out that this is possible. They are simply mapped onto
the Wronskians $W_{k_1,k_2}$ with $|k_1-k_2|>2$ \cite{DTb}, and they therefore
correspond to the other eigenvalue problems that were mentioned at the end
of section~1 above. Truncation of the fusion hierarchy can then be 
reinterpreted in terms of the (quasi-)periodicity 
(in $k$) that the functions $y_k$
exhibit whenever $M$ is rational. In the simplest cases 
(with $M$ rational and
$l(l{+}1){=}0$) this periodicity arises because the solutions to the
ODE live on a finite cover of $\CC\backslash\{0\}$; for other cases,
the monodromy around $x{=}0$ needs a little more care, but the story remains
essentially the same.

All good correspondences need a dictionary, and to end this section we
give a summary of the mapping between objects seen by the integrable model
and the Schr\"odinger equation:

\smallskip
\newcommand{\ppbox}[1]{\parbox{3.3cm}{\small #1}}
\newcommand{\pppbox}[1]{\,\parbox{2.7cm}{\small #1}}
\newcommand{\lra}{{}\!\!\!\!\leftrightarrow{~}}
\[
\begin{array}{|lcl|}
\hline
& & \\[-9pt]
{}~~\parbox{3cm}{Integrable\\[-1pt] Model}&
&{}~\parbox{3.3cm}{Schr\"odinger\\[-1pt] equation}\\[10pt]
\hline
& & \\[-10pt]
\pppbox{Spectral parameter}&\lra&\ppbox{Energy}\\[5pt]
\pppbox{Anisotropy}&\lra&\ppbox{Degree of potential}\\[5pt]
\pppbox{Twist parameter}&\lra&\ppbox{Angular momentum}\\[7pt]
\pppbox{(Fused) transfer\\ matrices}&\lra&\ppbox{Spectral
problems defined at $|x|{=}\infty$}\\[14pt]
\pppbox{Q operators}&\lra&\ppbox{
Spectral problems linking $|x|{=}\infty$ and $|x|{=}0$}\\[14pt]
\pppbox{Truncation of the\\ fusion hierarchy}&\lra
&\ppbox{Solutions on finite covers of $\CC \backslash
\{0\}$}\\[10pt]
\hline
\end{array}
\]

\noindent
(The two classes of spectral problems mentioned in this table
are related 
to the `lateral connection' and `radial connection' problems
in general WKB theory -- see, for example, \cite{OLV}.)

Armed with the dictionary,  the horizontal axis of figure \ref{fig1}
can be annotated
to indicate which integrable models correspond to the various values
of $N$ in the Bender-Boettcher problem. Thus for $N=1,2,3,4$ and $6$,
the relevant
integrable models are the $N{=}2$ SUSY point of the sine-Gordon model, 
the free-fermion
point, the Yang-Lee model, $\ZZ_4$ parafermions and the $4$-state Potts 
model respectively.
It is amusing that the $x^3$ potential is related by the 
correspondence to the Yang-Lee model (or, strictly speaking, to the sine-Gordon
model at the value of the coupling which allows for a reduction to Yang-Lee),
thus returning by a very indirect route to a neighbourhood
of the original thought of Bessis and Zinn-Justin.

\section{Generalisations}
The Bethe ansatz equations seen so far can all be written in terms of the
variable $E$ as
\[
\prod_{j=1}^{\infty}\left(\frac{E_j-\omega^{2M}E_k}{E_j-\omega^{-2M}E_k}%
\right)=-\omega^{2l{+}1},\quad k=1,2\dots
\]
where $\omega=e^{2\pi i/(2M{+}2)}$, $M$ is
related to the quantum group deformation
parameter, or 
anisotropy, of the lattice model, and $l$ is related to the twist.

These are the $n{=}2$ cases of a general family of $SU(n)$-related 
Bethe ansatz systems, relating $n{-}1$ sets of unknowns $\{E^{(m)}_k\}$,
with $m=1,2\dots n{-}1$ and $k=1,2\dots\infty$\,:
\[
\prod_{t=1}^{n{-}1}
\prod_{j=1}^{\infty}\left(\frac{E^{(t)}_j
-\omega^{\frac{nM}{2}C_{mt}}E^{(m)}_k}%
{E^{(t)}_j-\omega^{-\frac{nM}{2}C_{mt}}E^{(m)}_k}%
\right)=-\omega^{n\tau_m{+}1}.
\]
As in the $SU(2)$ case,
$M$ can be viewed as a deformation parameter, but this time there are not
one but $n{-}1$ independent twists, $\tau_1$, $\tau_2$,\,\dots 
$\tau_{n{-}1}$. 
The indices
$m$ and $t$ should be thought of as living on an $SU(n)$ Dynkin diagram, of
which $C_{mt}$ is the Cartan matrix. 
To obtain these equations using operators defined directly
in a continuum quantum
field theory, as 
achieved in \cite{BLZ1,BLZ2} for the $SU(2)$ case,
appears to be a largely open problem, though the first steps have been 
undertaken in \cite{FRS}. But even without this motivation, it is very natural
to ask whether 
the correspondence described above can be extended to cover BA systems of these
more general types.

The answer is yes~\cite{DDT}, and it turns out that one has to turn
to higher-order ordinary differential equations. 
Earlier but less complete results in this direction
were obtained in \cite{DTc,Sb}; 
aspects of the problem are also discussed in the recent article 
\cite{Srev}. One of the main difficulties is to find a parametrisation
of the higher-order differential operators which incorporates
the twists in a manageable way. The solution found in \cite{DDT}
starts by defining an elementary first-order differential operator, $D(g)$:
\[
D(g)=\left(\frac{d}{dx}-\frac{g}{x}\right)~.
\]
Elementary properties are $D(g)^{\dagger}=-D(-g)$ and 
$D(g_2{-}1)D(g_1)=D(g_1{-}1)D(g_2)$. Now, given a vector ${\bf g}=
(g_0,g_1\dots g_{n-1})\,$, set
\bea
&&
\!\!\!\!\!\!\!D({\bf g})=\nn\\
&&
\quad\,
D(g\phup_{n-1}{-}(n{-}1))
D(g\phup_{n-2}{-}(n{-}2))
\dots
D(g\phup_0)\nn
\eea
and impose $\sum_{i=0}^{n-1}g_i=n(n{-}1)/2$ 
to ensure that the $(n{-}1)^{\rm th}$
order derivative term vanishes (this allows
various theorems about Wronskians to hold in their simplest forms).
With this notation in place,
the ODE to consider is an immediate generalisation
of those seen earlier:
\[
\left( (-1)^{n+1}D({\bf g})+P(x,E)\right)\psi(x)=0
\]
with $P(x,E)=x^{nM}-E$. 
After some work, it turns out
this ODE does indeed contain a hidden
set of
$SU(n)$ Bethe ansatz equations. The parameter $M$
in $P(x,E)$
is equal to the $M$ appearing in the $SU(n)$ BAE quoted above, 
while 
the vector of parameters ${\bf g}$ is related to 
the twists in the BAE by
\[
{\bf g}
=
\left(
\begin{array}{c}
0\\ 1\\ 2\\ \vdots \\ \! n{-}1 \!
\end{array}
\right)
+
\left(
\begin{array}{cccc}
1{-}n & \, 2{-}n \, & 3{-}n & \dots \\ 
1 & 2{-}n & 3{-}n & \dots \\ 
1 & 2 & 3{-}n & \dots \\ 
\vdots & \vdots & \vdots &  \\ 
1 & 2 & 3 & \dots 
\end{array}
\right)\!\!
\left(
\begin{array}{c}
\tau\phup_1\\ \tau\phup_2\\ \vdots \\ \tau\phup_{\!n{-}1} 
\end{array}
\right)
\]
The $SU(n)$ structure
is encoded via certain Wronskians
$W[y_i]$,
$W[y_i,y_j]$, $W[y_i,y_j,y_k]$\,\dots\
These are $m$-dimensional determinants of matrices formed by the functions 
$y\phup_{k_1}$,\dots
$y\phup_{k_m}$ and their first $(m{-}1)$ derivatives, for $m=1\dots n{-}1$.
The functions $y_k$ themselves are certain special solutions of the ODE,
subdominant in particular sectors of the
complex plane. 
They generalise the
$y_k$ introduced by Sibuya for second-order ODEs, that were
described in
section~3 above.
For the precise definitions and more details of how the mapping goes, the paper 
\cite{DDT} should be 
consulted, since space prevents a fuller discussion in this short review.

\section{Conclusions}
The headline conclusion of this talk should already be clear: it
is that the $T$ and $Q$ operators
which arise in certain integrable quantum field theories
encode spectral data, at least in their ground-state eigenvalues. 
This gives a novel perspective
on the Bethe ansatz, and also a new way to treat spectral problems via
the solution of nonlinear integral equations. 

One important topic not
covered here is the new light that the correspondence
sheds on some previously-conjectured duality 
properties of integrable models \cite{BLZa,DTb,DTc,DDT}. 
A generalisation of the Langer \cite{La} transformation 
can be employed to map the ODE 
with the potential $x^M$ to one with potential
$x^{\tilde M}$, $\tilde M=-M/(M{+}1)$.
For $n=2$, this sends the parameter
$q=e^{i\pi/(M{+}1)}$ to $\widetilde q=
e^{i\pi/(\tilde M{+}1)}=
e^{i\pi(M{+}1)}$,
which is precisely the kind of
duality discussed in the talk by F.~Smirnov at
this conference.

There are many further problems to be explored, 
of which we list just a few. First,
one would like to know how many other BA systems can be
brought into the correspondence, beyond the $A_{n-1}$-related cases
described above, and whether more general polynomial potentials might
also have a role to play. The set is certainly not empty --
see \cite{Sc} -- but the problem of finding ODEs even for the $D$ and
$E$ related BA systems
remains open. Second, the correspondences established to date have
all concerned massless integrable lattice models, in a `field theory'
limit where the number of 
sites, and of Bethe ansatz roots, tends to infinity.
Correspondences for more general massive models, and for lattice models
with a finite number of sites, would be very interesting. Finally, we should
admit that our observations remain at a rather formal
and mathematical level. At some stage one should ask what physics lies behind
all of this, but perhaps such questions will have to
wait until the answers to the other open problems have
been found and classified. In this sense,
we may still be in a `stamp-collecting'
phase, and we can expect that further work
will lead to a much more systematic
understanding of the whole story.

\acknowledgments
%
We would like to 
thank Junji Suzuki and Andr\'e Voros for their
helpful comments.
PED thanks the organizers of the TMR2000 meeting for the
invitation to speak.
We were supported in part by a TMR grant of the
European Commission, reference ERBFMRXCT960012.
%
%
%

\end{document}